\documentclass[a4paper]{IEEEtran}

\pdfoutput=1

\usepackage{amssymb}
\usepackage{float}
\usepackage{psfrag}
\usepackage{graphicx}
\usepackage{cite}
\usepackage{cooltooltips}
\usepackage[centertags]{amsmath}
\usepackage{dsfont}
\usepackage{color}

\newcommand{\revd}[1]{{}}                
\newcommand{\revv}[1]{{#1}}                
\newcommand{\rev}[1]{{#1}}                
\newcommand\vect[1]{\mathbf #1}
\newcommand{\field}[1]{\mathbb{#1}}                
\newcommand{\R}{\field{R}}                         
\newcommand{\C}{\field{C}}                         
\newcommand{\Z}{\field{Z}}                         
                      
\renewcommand{\H}{\mathds{H}} 
\newcommand{\D}{\mathds{D}} 
\newcommand{\M}{\mathds{M}} 
\renewcommand{\S}{\mathds{S}} 
\newcommand{\Hm}{\mathbf{H}} 
\newcommand{\Dm}{\mathbf{D}} 
\newcommand{\Mm}{\mathbf{M}} 
\newcommand{\A}{\rev{\mathbf{X}}}
\newcommand{\s}{\mathbf{s}}

\newcommand{\ii}{i}

\newcommand{\rem}[1]{}

\newcommand{\ist}{\hspace*{.1mm}}
\newcommand{\rmv}{\hspace*{-.1mm}}

\makeatletter
\renewcommand\fs@boxed{\def\@fs@cfont{\bfseries}\let\@fs@capt\floatc@plain \def\@fs@pre{\setbox\@currbox\vbox{\hbadness10000
\moveleft3.4pt\vbox{\advance\hsize by6.8pt \hrule \hbox to\hsize{\vrule\kern3pt
\vbox{\kern3pt\box\@currbox\kern3pt}\kern3pt\vrule}\hrule}}}%
\def\@fs@post{\vspace*{-2mm}}\let\@fs@iftopcapt\iffalse}
\makeatother

\floatstyle{boxed}
\newfloat{floatbox}{thp}{lob}[section]
\floatname{floatbox}{Text Box}

\begin{document}

\title{  
~\\[-10mm]
\cooltooltiptoggle{Time-Frequency Foundations of Communications}\vspace{2mm}}
\author{%
\authorblockN{\em Gerald Matz, 
Helmut B\"olcskei,
and Franz Hlawatsch
\parbox{0.9\textwidth}{%
 ~\\ \begin{quote}\small
Hitherto communication theory was based on two alternative methods of signal analysis.	One is the description of the signal as a function of time; the other is Fourier analysis.	Both are idealizations, as the first method operates with sharply defined instants of time, the second with infinite wave-trains of rigorously defined frequencies. But our everyday experiences---especially our auditory sensations---insist on a description in terms of both time and frequency. \,--- 
{\em Dennis Gabor \cite{Gabor46a}}
\end{quote}
\vspace*{-15mm}
\vspace{8mm}}
}}

\maketitle

\section{Introduction and Background}

In the tradition of Gabor's 1946 landmark paper \cite{Gabor46a}, we advocate a time-frequency (TF) approach to communications. 
TF methods for communications have been proposed very early (see the box \textsc{History}).
While several tutorial papers and book chapters on the topic are available
(see, e.g., \cite{wanggg00,magg06,ltvbook11} and references therein), the goal of this paper is to present the fundamental
aspects in a coherent and easily accessible manner. Specifically, we establish the role of TF methods in communications across a range of 
subject areas including TF dispersive channels, orthogonal frequency division multiplexing (OFDM), 
information-theoretic limits, and system identification and channel estimation. Furthermore, we present fundamental results that are stated in the literature 
for the continuous-time case in simple linear algebra terms.

\begin{figure*}[tbh]
\renewcommand\baselinestretch{1.1}\small\normalsize
\begin{center}
\noindent\includegraphics[width=0.8\textwidth]{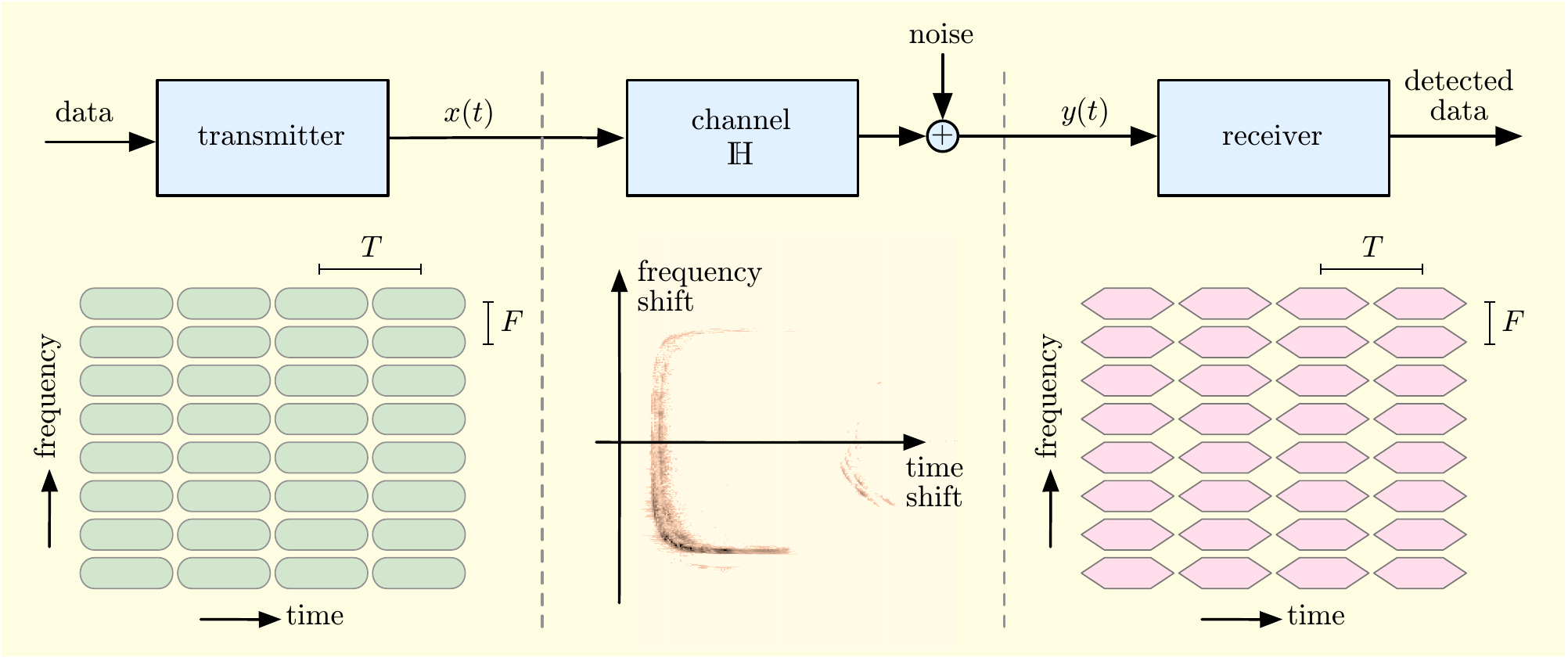}
\end{center}
\vspace*{-5mm}
\caption{\label{fig:intro}Top: A communication system consisting of a transmitter, a noisy channel, and a receiver. 
Bottom: Illustration of the TF shift structure of an OFDM modulator, a (measured) TF dispersive channel (see \cite{cfm_proc11}), and an OFDM demodulator.}
\vspace*{-1mm}
\end{figure*}

We consider a point-to-point communication scenario with a single transmitter, a channel, and a single receiver as shown in Fig.~\ref{fig:intro}.
The channel models the transmission medium and imperfections of transmitter and receiver hardware like oscillators, amplifiers, and antennas.

A basic element of TF analysis is the TF shift operator $\M_\nu\D_\tau$, which induces a delay (time shift) $\tau$ and a modulation 
\revv{(frequency shift)}
$\nu$ according to 
$\big(\M_\nu\D_\tau x)(t) = x(t-\tau)e^{j2\pi\nu t }$ \cite{groebook,fla-book2}. TF shifts are fundamental to communications in a twofold manner:

\begin{enumerate}

\item Many linear channels are TF dispersive, i.e., they induce time dispersion (delays) and frequency dispersion (modulation).
These channels can be represented as a weighted superposition of TF shift operators \cite{groebook}.

\vspace{.5mm}

\item OFDM is a multicarrier transmission scheme that modulates the data symbols onto Weyl-Heisenberg (WH) function sets, 
also known as Gabor sets \cite{groebook,Fei_Str_97}. These function sets consist of TF shifted versions of a prototype pulse (Gabor's ``logons'' \cite{Gabor46a}). 

\vspace{.5mm}

\end{enumerate}

\begin{floatbox}[t!]
\vspace*{1mm}
\begin{center}
\begin{minipage}[t]{0.98\columnwidth}

\begin{center}
{\sc History}  
\end{center}
\vspace*{-1mm}

TF analysis has been linked with communications for a long time. Gabor \cite{Gabor46a}, 
father of the Gabor expansion, proposed the use of ``TF logons'' (TF shifts of a prototype pulse) to represent communication signals.
Zadeh \cite{Zad50} introduced a TF transfer function of TF dispersive systems.
Chang \cite{Chang66} proposed the multicarrier transmission scheme known as orthogonal frequency division multiplexing (OFDM).
Kailath \cite{Kai62} discussed the sampling and measurement of TF dispersive systems. 
Bello \cite{Bel63} studied random TF dispersive channels and introduced the concept of wide-sense stationary uncorrelated scattering (WSSUS). 
The estimation of WSSUS channel statistics was addressed by Gallager \cite{Gal64} and Gaarder \cite{Gaa68}. An extensive
discussion of communication over random TF dispersive channels was provided by Kennedy \cite{Ken69}. 

\end{minipage}
\end{center}
\vspace*{-1mm}
\end{floatbox}

OFDM and TF dispersive channels are at the heart of a broad range of communication systems,  including
Digital Audio/Video Broadcasting, Wireless Local Area Networks (IEEE 802.11), Wireless Metropolitan Area Networks (IEEE 802.16), 
3GPP Long-Term Evolution, Wireless Personal Area Networks (e.g., WiMedia), Vehicular Ad Hoc Networks, L-band Digital Aeronautical Communication Systems, 
Digital Subscriber Lines, power-line communications, and underwater acoustic communications
\cite{cfm_proc11,ladebusch06,hanzo-mimo-ofdm,haas2002,corripio06,baoshengli08,antonia_book11}. 
The purpose of this paper is to discuss the relevance of TF analysis to OFDM and TF dispersive channels, and to demonstrate that 
WH frame theory 
\cite{christensen03}
and TF operator representations are powerful tools for pulse design \cite{Haas96,kozmol98,boel_spie99,schniter_tsp07,Matz-Charlypaper07},
capacity analysis \cite{durisiboel_book11}, and channel identification (sounding, estimation) \cite{kozek_identification_2005,sounder_tc00,heckelboel12}. 
We note that parts of this paper draw on our previous work in 
\cite{GM-Hla-jmp98,boel_spie99,Matz-Charlypaper07,heckelboel12,durisiboel_book11}, and \cite{ltvbook11_ch1}.

\section{TF Dispersive Channels}

In this section, we discuss the physics, system theory, and statistics of TF dispersive channels.

\vspace*{-2mm}

\subsection{Physics} 

First, we describe various physical mechanisms that entail a superposition of TF shifts.

\vspace*{1mm}

\def\sol{c}

\subsubsection{Multipath Propagation and Doppler Effect} In wireless (radio or underwater) communications, the electromagnetic or acoustic wave 
propagating from the transmitter to the receiver may interact with objects in the environment. These objects are commonly
referred to as scatterers, even though the interaction mechanism may include reflection and diffraction. The wave usually propagates along several 
distinct paths with different propagation delays and attenuation factors. This situation is known as multipath propagation. 

If transmitter, receiver, or scatterers are  moving, the 
Doppler effect entails a time scaling (equivalently, a frequency scaling)
of multipath components. For narrowband signals, i.e., signals whose spectrum is supported in a small band around a carrier frequency $f_\text{c}$, 
frequency scaling can be approximated by a frequency shift of $\nu = v f_\text{c}/\sol$, where $\sol$ is the wave propagation speed and 
$v$ is the velocity of the moving object in the direction of wave propagation. 

In general, the transmitted signal $x(t)$ is affected by both multipath propagation and Doppler frequency shifts. Assuming $I$ specular paths
with delays $\tau_i$, Doppler frequencies $\nu_i$, and complex gains $S_i$, the receive signal is given by the following 
weighted superposition of TF shifts of $x(t)$:\footnote{Additive noise is neglected throughout this section.}
\[
y(t) 
= \sum_{i=1}^I S_i\,x(t-\tau_i)\,e^{j2\pi\nu_i t} = \sum_{i=1}^I S_i\ist\big(\M_{\nu_i}\D_{\tau_i} x)(t).
\vspace*{1mm}
\]

\subsubsection{Medium Variations} Many transmission media, such as cables and optical fibers, are characterized by 
material dispersion, i.e., a group velocity that varies with frequency. Material dispersion can be modeled by a time-dispersive channel that is 
described by the convolution relation\footnote{Integrals are over the entire real line.}
$y(t)= \int g(\tau)\,x(t-\tau)\,d\tau = \int g(\tau)(\D_\tau x)(t)\,d\tau$, i.e., the receive signal is a weighted superposition of time-shifted versions of the 
transmit signal.

In the presence of environmental changes, switching effects, or component drift, the transmission medium varies over time. 
Such variations can be modeled by a frequency-dispersive channel with multiplicative input-output relation $y(t)=m(t)x(t)$.
Denoting the Fourier transform of $m(t)$ by $M(\nu)$, the equivalent relation $y(t)= \int M(\nu)(\M_\nu x)(t)\,d\nu$ shows that the
receive signal is a weighted superposition of frequency-shifted versions of the transmit signal. 
General channels may exhibit both time and frequency dispersion.

\vspace*{1mm}

\subsubsection{Oscillator Imperfections and Timing Offsets} In most communication systems, the baseband transmit signal is modulated onto a 
sinusoidal carrier via an oscillator. The receive signal is then demodulated, ideally using the same sinusoidal carrier. 
However, practical oscillators exhibit imperfections such as frequency offset and phase noise. Furthermore, transmitter and receiver suffer from a timing (clock) offset.
Consider, for example, a receiver with frequency offset $\Delta f$, phase noise $\phi(t)$, and timing offset $\Delta t$, and an otherwise ideal transmission medium.
Here, the baseband receive signal is given by 
\[
y(t) = e^{-j2\pi f_\text{c} \Delta t}\! \int \Psi(\nu\!+\!\Delta f) (\M_\nu\D_{\Delta t} x)(t)\,d\nu ,
\]
where $\Psi(\nu)$ denotes the Fourier transform of $e^{-j2\pi \phi(t)}$. Frequency offset, phase noise, and timing offset thus amount to a superposition of TF shifts.
	
\vspace{-1mm}

\subsection{Elementary Channel Characterizations} 

We next review system-theoretic aspects of TF dispersive channels. In what follows, frequency shifts will 
be referred to as Doppler shifts even if the underlying physical mechanism is not the Doppler effect.
The basic input-output relation of a TF dispersive channel $\H$ is denoted as $y(t)=(\H x)(t)$.

\vspace*{1mm}

\subsubsection{Delay-Doppler Spreading Function} 

We have seen that different physical effects amount to a weighted superposition of TF shifts. In fact, it is shown in \cite[Th.~14.3.5]{groebook} 
that virtually any linear channel (operator) $\H$ can be represented as a (generally continuous) superposition of TF shift operators in the sense that
\begin{equation}
y(t) = \iint  S_{\H}(\tau,\nu)\, \big(\M_\nu\D_\tau x)(t) \,d\tau\,d\nu.
\label{eq:spreading_io}
\end{equation}
For the finite-dimensional case, a simple explanation of this representation result is given in the box {\sc Operator Representation}.
The function $S_{\H}(\tau,\nu)$ in \eqref{eq:spreading_io} characterizes the complex weight associated with delay $\tau$ and Doppler shift $\nu$ and is known
as delay-Doppler spreading function. We note that even though \eqref{eq:spreading_io} applies generally, in the (ultra)wideband regime more parsimonious
channel representations may be obtained using Doppler scaling instead of Doppler shifts \cite{antonia_book11}.

\begin{floatbox}[t!]
\vspace*{1mm}
\begin{center}
\begin{minipage}[t]{0.98\columnwidth}

\begin{center}
{\sc Operator Representation}  
\end{center}
\vspace*{-1mm}

The input-output relation (\ref{eq:spreading_io}) describes a large class of linear operators.  This can easily be proved for a finite-dimensional setting using basic linear algebra. 
We define the $N\times N$ cyclic time-shift matrix ${\bf D}$, which has ones in the \rev{subdiagonal} and in the top right corner and zero entries else,
and the diagonal $N\,\times\,N$ modulation (frequency shift) matrix ${\bf M}$, which has $e^{-j2\pi (i-1)/N}\!$, $i \in \{1,...,N\},$ as its $i$th main diagonal entry. 
The inner product on $\C^{N\,\times\,N}$ is defined as
$
\langle{\bf A},{\bf B}\rangle=\mbox{Tr}\,({\bf B}^{H}\!{\bf A}),
$
where $\mbox{Tr}\,(\cdot)$ denotes the trace and the superscript $^H$ stands for Hermitian transposition.
It can be shown that the $N^2$ matrices $\big\{ \frac{1}{\sqrt{N}}{\bf M}^{l}{\bf D}^{m}\big\}_{m,l \in \{0,...,N-1\}}$ 
form a complete orthonormal set for $\C^{N\times N}$. Hence, every ${\bf H}\,\in\,\C^{N\times N}$ can be decomposed as
\begin{equation}\label{eq:sfdiscrete}
\Hm  = \sum_{m=0}^{N-1} \sum_{l=0}^{N-1} S_{\Hm}[m,l]\, \Mm^l\Dm^m 
\end{equation}
with $S_{\Hm}[m,l] = \langle \Hm, {\bf M}^{l}{\bf D}^{m} \rangle/N$. This is the discrete, finite-dimensional counterpart of \eqref{eq:spreading_io}.

\end{minipage}
\end{center}
\vspace*{-1mm}
\end{floatbox}

\vspace{1mm}

\subsubsection{Channel Spread and Underspread Property}
Most channels are {underspread}, i.e., the amount of delay-Doppler spreading they induce is small in that their spreading function $S_{\H}(\tau,\nu)$ 
is effectively confined to a small region in the delay-Doppler plane. An example is visualized in the bottom center plot in Fig.~\ref{fig:intro}. 
Selected aspects of the underspread property are considered in the box \textsc{Underspread Channels}, again in a finite-dimensional setting.

For spreading functions with finite support, a formal definition of the underspread property 
can be obtained by circumscribing the support region with a rectangle that is centered around the 
origin\footnote{The center of the rectangle is immaterial for the definition of the underspread property and is chosen to be the origin 
for simplicity of exposition.}
and whose side lengths equal twice the channel's maximum delay $\tau_{\max}$ and maximum Doppler frequency $\nu_{\max}$, respectively.
The area of this rectangle, $d_\H = 4\tau_{\max}\nu_{\max}$, measures the channel's overall TF dispersion and is referred to as the channel spread. 
A channel is then said to be underspread if $d_\H \le 1$ and overspread if $d_\H > 1$.  For spreading functions that do not have finite support,
the channel spread can be quantified in terms of moments \cite{GM-Hla-jmp98}. 

For multipath propagation we have $d_\H \propto 1/\sol^2$ \cite{ltvbook11_ch1}. Hence, the channel spread of radio channels 
(where $\sol$ equals the speed of light) is typically much smaller than that of underwater acoustic channels (where $\sol$ equals the speed of sound). 
In fact, radio channels have $d_\H$ on the order of $10^{-6}$ to $10^{-3}$ and thus are highly underspread, whereas underwater acoustic channels 
can even be overspread.

\begin{floatbox}[t!]
\vspace*{1mm}
\begin{center}
\begin{minipage}[t]{0.98\columnwidth}

\begin{center}
{\sc Underspread Channels}  
\end{center}
\vspace*{-1mm}

Consider a finite-dimensional channel $\Hm \in \mathbb{C}^{N\times N}\!$, and assume that the discrete spreading function $S_\Hm[m,l]$ 
in \eqref{eq:sfdiscrete} is supported on a small set $\mathcal{S}$ around the origin, i.e., $S_\Hm[m,l]=0$ for $(m,l)\not\in\mathcal{S}$.
The sum in \eqref{eq:sfdiscrete} then consists of only $|\mathcal{S}|$ nonzero terms. The channel is underspread if $|\mathcal{S}|\le N$, 
i.e., the number $|\mathcal{S}|$ of degrees of freedom of the channel does not exceed the dimensionality $N$ of the 
ambient signal space.

A key observation explaining many properties of underspread channels is the fact that for small $m$ and small $l$, 
time shifts $\Dm^{m}$ and frequency shifts $\Mm^{l}$ commute approximately. Specifically, using the (non-)commutation relation 
$\Mm^{l}\Dm^{m} = \Dm^{m}\Mm^{l}e^{j2\pi\frac{ml}{N}}\rmv$ and the bound $|1\rmv-\rmv e^{j2\pi\phi}| \le 2\pi|\phi|$, one obtains 
\begin{equation}\label{eq:TFcommute}
\|\Dm^{m}\Mm^{l} - \Mm^{l}\Dm^{m} \| \le 2\pi\frac{|ml|}{N}  \|\Dm^{m}\Mm^{l}\| ,
\end{equation}
where $\|\cdot\|$ is an arbitrary matrix norm. Clearly, if $|ml|$ is small relative to $N$, \eqref{eq:TFcommute} implies $\Dm^{m}\Mm^{l} \approx \Mm^{l}\Dm^{m}$.
\revv{In combination with \eqref{eq:sfdiscrete}, this approximate commutation property implies}
that underspread channels commute approximately.

We next demonstrate the approximate multiplicativity property \eqref{eq:appmult} in the finite-dimensional setting.
Here, the discrete TF transfer function $L_\Hm[n,k]$ equals the discrete two-dimensional Fourier transform of the discrete spreading function $S_\Hm[m,l]$, 
i.e., $L_\Hm[n,k] = \sum_{m=0}^{N-1}\sum_{l=0}^{N-1} S_\Hm[m,l] e^{-j2\pi\frac{km-nl}{N}}$ (cf.~\eqref{eq:transspreadrel}).
For underspread channels $\Hm_1$ and $\Hm_2$ with identical spreading function support $\mathcal{S}$, 
the approximation $L_{\Hm_1\Hm_2}[n,k]\approx L_{\Hm_1}[n,k]L_{\Hm_2}[n,k]$ (cf.\ \eqref{eq:appmult}) translates into the approximate convolution
\begin{equation}\label{eq:TFconv}
S_{\Hm_1\Hm_2}[m,l] \approx \! \sum_{(m',l')\in\mathcal{S}} 
S_{\Hm_1}[m\rmv-\rmv m',l\rmv-\rmv l'] S_{\Hm_2}[m',l'] .
\end{equation}
To prove \eqref{eq:TFconv}, we start with the expression $S_{\Hm_1\Hm_2}[m,l] = \langle \Hm_1\Hm_2, {\bf M}^{l}{\bf D}^{m} \rangle/N$
and replace $\Hm_1$ and $\Hm_2$ by their spreading representations \eqref{eq:sfdiscrete}. This yields
\begin{multline}
\!\!\!S_{\Hm_1\Hm_2}[m,l] = \! \sum_{(m'',\,l'')\in\mathcal{S}} \! S_{\Hm_1}[m'',l''] \sum_{(m',\,l')\in\mathcal{S}} \! S_{\Hm_2}[m',l']\\  
\cdot \langle \Mm^{l''}\Dm^{m''}\Mm^{l'}\Dm^{m'} \!, {\bf M}^{l}{\bf D}^{m} \rangle / N .
\label{eq:S_H_12}
\end{multline}
Now, thanks to \eqref{eq:TFcommute} and the orthogonality of the TF shift matrices ${\bf M}^{l}{\bf D}^{m}$, we obtain 
$\langle \Mm^{l''}\Dm^{m''}\Mm^{l'}\Dm^{m'}, {\bf M}^{l}{\bf D}^{m} \rangle/N \!\approx\! \delta_{(l-l'-l'') \,{\rm mod}\, N} \, \delta_{(m-m'-m'') \,{\rm mod}\, N}$.
Inserting this into \eqref{eq:S_H_12} yields \eqref{eq:TFconv}.

\end{minipage}
\end{center}
\vspace{-1mm}
\end{floatbox}

\vspace{1.5mm}

\subsubsection{TF Transfer Function} 
The spreading function represents channels in the delay-Doppler domain. A dual TF representation, termed {TF transfer function}, represents
channels in the TF domain and is defined as the two-dimensional (2-D) Fourier transform of the spreading function \cite{Zad50,ltvbook11_ch1,GM-Hla-jmp98}:
\begin{equation}\label{eq:transspreadrel}
L_\H(t,f) = \iint S_\H(\tau,\nu) \, e^{-j2\pi(f\tau-t\nu)}\,d\tau\,d\nu.
\end{equation}
For underspread channels, $L_\H(t,f)$ is smooth and characterizes the channel's TF weighting behavior. 
This generalizes the frequency response $H(f)$ of time-invariant channels.

The complex exponentials $x(t) = e^{j2\pi f_0 t}$, $f_0 \!\in\! \mathbb{R}$, are eigen\-functions of all linear time-invariant channels.
For TF dispersive channels, a universal set of structured eigenfunctions does not exist. Underspread channels, however, 
satisfy the approximate eigenrelation
\begin{equation}\label{eq:appeigen}
\big(\H g_{t_0,f_0}\big)(t) \approx L_\H(t_0,f_0) \, g_{t_0,f_0}(t) ,
\end{equation}
with $g_{t_0,f_0}(t) = \big( \M_{f_0}\D_{t_0}g \big)(t)$; the accuracy of \eqref{eq:appeigen} depends on how well the function $g(t)$ is localized 
(around time $0$ and frequency $0$). Thus, $L_\H(t_0,f_0)$ is the approximate eigenvalue associated with an approximate eigenfunction that is 
TF localized around the TF point $(t_0,f_0)$.
This property entails an approximate diagonalization of the channel and explains why OFDM is a natural choice for signaling over 
underspread TF dispersive channels.

For underspread channels, the TF transfer function is furthermore approximately multiplicative, i.e.,
\begin{equation}\label{eq:appmult}
L_{\H_1\H_2}(t,f) \approx L_{\H_1}(t,f) L_{\H_2}(t,f) . 
\end{equation}
This implies that underspread channels commute approximately, i.e., $\H_1\H_2\approx\H_2\H_1$.
The approximate commutation of underspread channels  is of practical importance, e.g., in channel sounding \cite{sounder_tc00}.
A derivation of \eqref{eq:appmult} in the finite-dimensional setting is given in the box \textsc{Underspread Channels}.

The approximations \eqref{eq:appeigen} and \eqref{eq:appmult} nicely show that, in terms of transfer function calculus, 
underspread TF dispersive channels behave approximately like time-invariant channels. This is due to the fact that underspread channels 
share a structured set of approximate eigenfunctions.

\begin{figure}[t]
\renewcommand\baselinestretch{1.1}\small\normalsize
\begin{center}
\noindent\includegraphics[width=\columnwidth]{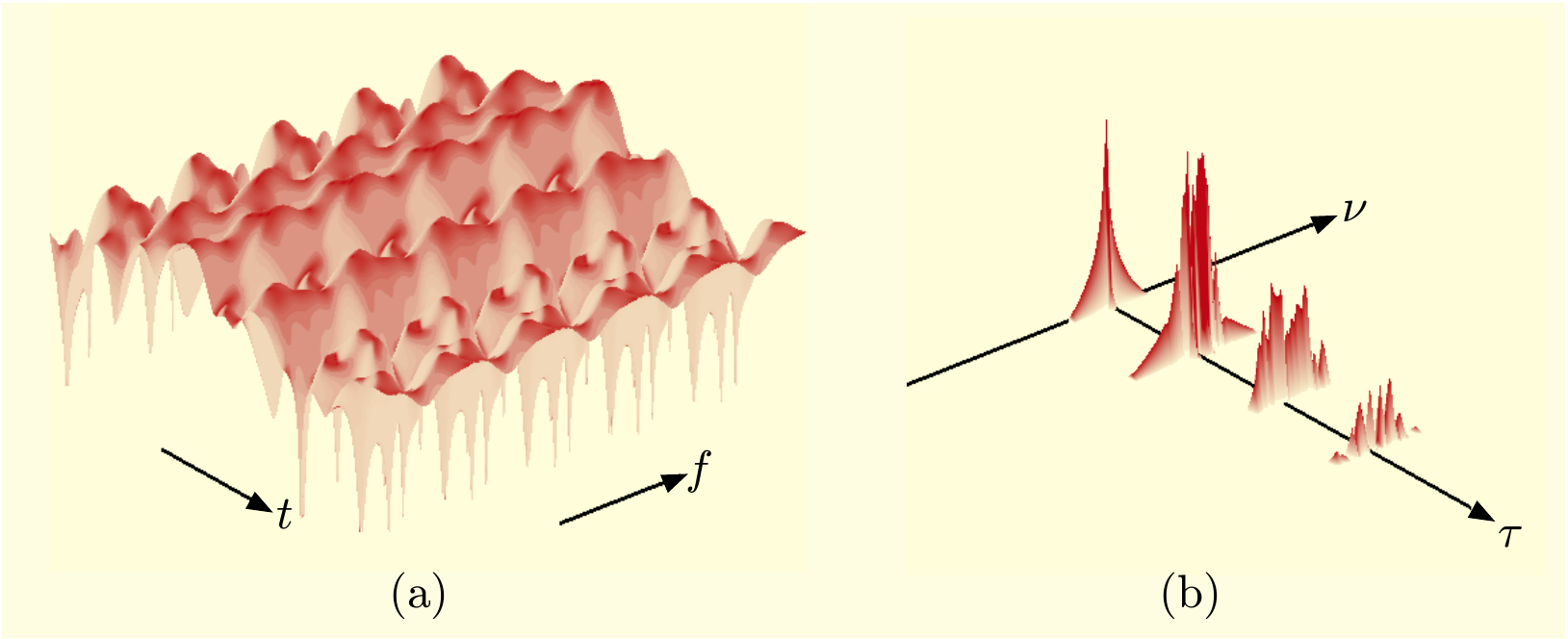}
\end{center}
\vspace*{-5mm}
\caption{\label{fig:channel_reps}
Example of an underspread TF dispersive channel with maximum delay $6$\,ms and Doppler spread $14.4$\,Hz:
(a) TF transfer function over a duration of $1$\,s and a bandwidth of $2.5$\,kHz,
(b) spreading function in the delay-Doppler region $[0,8)$\,ms$\times[-40,40)$\,Hz 
(outside this region, the spreading function is 40\,dB below the maximum value).
For both representations, the magnitude is displayed on a log scale with 40\,dB dynamic range.}
\vspace*{-2mm}
\end{figure}

In Fig.~\ref{fig:channel_reps}, we show the TF transfer function and spreading function of a realization of Channel 6 specified in the DRM standard \cite{etsi_drm}. 
This is an underspread channel that models skywave propagation. The echoes visible in Fig.~\ref{fig:channel_reps}(b) 
correspond to multiple reflections at the ionosphere.

\vspace{-1mm}

\subsection{Channel Statistics} 

Many channels are modeled as random; examples of the underlying phenomena include fading, unknown time and frequency offsets, and phase noise. 
The system functions  $S_\H(\tau,\nu)$ and $L_{\H}(t,f)$ then become 2-D random processes, with 4-D correlation (or covariance) functions.

\vspace{1.5mm}

\subsubsection{WSSUS Channels and Scattering Function}

An important simplification of the channel statistics is obtained for channels that are wide-sense stationary with uncorrelated scattering (WSSUS). Here,
$L_\H(t,f)$ is a process that is stationary in time and frequency. Hence, its correlation function is independent of $t$ and $f$, i.e.,
\[
\mathsf{E}\big\{ L_\H(t,f)\,L_\H^*(t-\Delta t,f-\Delta f)  \big\} = R_\H(\Delta t, \Delta f),
\]
where $R_\H(\Delta t, \Delta f)$ is known as the channel's TF correlation function. Correspondingly, the scatterer reflectivities described by the 
spreading function are uncorrelated, i.e., 
\[
\mathsf{E}\big\{ S_\H(\tau,\nu)\,S_\H^*(\tau',\nu')  \big\} = C_\H(\tau,\nu)\,\delta(\tau-\tau')\,\delta(\nu-\nu').
\] 
Here, $C_\H(\tau,\nu)\ge 0$ describes the average intensity of scatterers with delay $\tau$ and Doppler shift $\nu$ and is referred to as 
scattering function \cite{Bel63,ltvbook11_ch1}. The scattering function and the TF correlation function are related via a 2-D Fourier transform:
\[
C_\H(\tau,\nu) = \iint R_\H(\Delta t, \Delta f)\,e^{-j2\pi(\nu\Delta t -\tau\Delta f)}\,d\Delta t\,d\Delta f.
\]
This shows that $C_\H(\tau,\nu)$ can be interpreted as the delay-Doppler domain power spectral density of $L_\H(t,f)$.

The scattering function relates the time-varying power spectra of the transmit and receive signals according to \cite{ltvbook11_ch1}
\[
P_y(t,f) = \iint C_\H(\tau,\nu)\,P_{\M_\nu\D_\tau x}(t,f)\,d\tau\,d\nu,
\]
where $P_\cdot(t,f)$ is an arbitrary type I TF energy spectrum \cite{GM_spectra02}. This relation is the statistical TF counterpart of \eqref{eq:spreading_io}; 
it amounts to a convolution, since $P_{\M_\nu\D_\tau x}(t,f) = P_{x}(t-\tau,f-\nu)$.

A WSSUS random channel is said to be underspread if its scattering function $C_\H(\tau,\nu)$ is effectively confined to a small delay-Doppler region 
(the spreading function of every channel realization then is confined to the same region). Wireless (radio) channels are underspread also in this stochastic sense.

\vspace{1.5mm}

\subsubsection{Non-WSSUS Channels and Local Scattering Function}
Recently, high-mobility applications like vehicular communications have spurred interest in non-WSSUS channels \cite{ltvbook11_ch1,cfm_proc11}.
For non-WSSUS channels, $L_\H(t,f)$ is a nonstationary random process and different scatterer contributions are correlated.
A generalization of the scattering function $C_\H(\tau,\nu)$ to non-WSSUS channels is provided by the local scattering function \cite{ltvbook11_ch1},
which equals the (generalized) Wigner-Ville spectrum \cite{fla-book2,GM_spectra02} of $L_\H(t,f)$. 
The local scattering function $\mathcal{C}_\H(t,f;\tau,\nu)$ describes the average power of scatterers that cause a delay $\tau$ and a Doppler shift $\nu$ 
of the transmit signal component localized around time $t$ and frequency $f$.

\vspace*{-1mm}

\section{OFDM}

\vspace*{.5mm}

In the spirit of Gabor \cite{Gabor46a}, OFDM transmits data symbols via TF logons (TF shifts of a prototype pulse).
OFDM is used in a large number of wireless and wireline communication systems and standards. 
Among other reasons, OFDM is popular because cyclic prefix (CP) OFDM diagonalizes time-invariant channels
and, more generally, well TF localized WH sets approximately diagonalize underspread TF dispersive channels. 
Here, we consider \emph{pulse-shaping} OFDM, which constitutes a unified framework for CP-OFDM \cite{Peled80}, zero-padded OFDM \cite{wanggg00}, 
DFT filterbank modulation \cite{FBMC_JASP10}, and, with a slight modification, OFDM with offset quadrature amplitude modulation (OQAM) \cite{boel_spie99}.

\vspace*{-1mm}

\subsection{Modulation and Demodulation} 

The transmit signal in a pulse-shaping OFDM system is formed by modulating data symbols $c_{n,k}$ onto TF shifted versions of a transmit pulse $g(t)$ 
(e.g., \cite{kozmol98,hanzo-mimo-ofdm,Matz-Charlypaper07,boel_spie99}):
\begin{equation}\label{eq:ofdmmod}
x(t) = \sum_{n\in\Z} \sum_{k\in\Z} c_{n,k}\,g_{n,k}(t), 
\vspace*{-2mm}
\end{equation}
with
\vspace*{-1mm}
\[
g_{n,k}(t) = \big( \M_{kF}\D_{nT} g \big)(t) = g(t-nT)e^{j2\pi kFt} . 
\vspace*{.5mm}
\]
Here, $T$ is the OFDM symbol duration and $F$ is the subcarrier spacing. (In practical OFDM systems, the sum with respect to $k$ involves only a finite number of subcarriers. 
We assume infinitely many subcarriers to simplify the presentation.) The collection of ``logons'' $\{g_{n,k}(t)\}_{n,k \in \mathbb{Z}}$ is known as a WH set. 
Its TF localization structure is schematically illustrated in the bottom left plot in Fig.~\ref{fig:intro} (in reality, the $g_{n,k}(t)$ overlap in time or in frequency).
To recover the data symbols $c_{n,k}$, the receiver projects the receive signal $y(t)$ onto TF shifted versions of a receive pulse $\gamma(t)$
by computing the inner products 
\begin{equation}\label{eq:ofdmdemod}
\hat{c}_{n,k}=\big<y,\gamma_{n,k}\big>=\int y(t)\,\gamma_{n,k}^*(t)\,dt, 
\end{equation}
with $\gamma_{n,k}(t) = \big( \M_{kF}\D_{nT} \gamma \big)(t) = \gamma(t-nT)e^{j2\pi kFt}$. This OFDM demodulation is followed by further receiver processing such as
channel estimation, demapping, and decoding.

In the absence of channel distortions and noise, 
\revv{it is desirable to have}
perfect symbol recovery, 
i.e., $\hat{c}_{n,k} \!=\! {c}_{n,k}$\revv{; this} is guaranteed if the WH sets  
$\{g_{n,k}(t)\}_{n,k \in \mathbb{Z}}$ and $\{\gamma_{n,k}(t)\}_{n,k \in \mathbb{Z}}$ satisfy the biorthogonality condition
$\langle g_{n,k},\gamma_{n^{\prime}\!,k^{\prime}} \rangle = \delta_{n-n^{\prime}}\delta_{k-k^{\prime}}$. Biorthogonality presupposes $TF \!>\! 1$, 
in which case the system is said to employ a TF guard region \cite{Haas96}. CP-OFDM \cite{Peled80} and zero-padded OFDM \cite{wanggg00}
are special cases, with the TF guard region being a temporal guard region only. However, $TF\!>\! 1$ can also be achieved by introducing a spectral guard region via an 
increase of the subcarrier spacing $F$; this can reduce intercarrier interference in frequency-dispersive environments.
The spectral efficiency of an OFDM system is inversely proportional to $TF$ and is thus determined by the density of the TF grid $\big\{(nT,kF)\big\}_{n,k\in\Z}$.

\vspace*{-1mm}

\subsection{Analysis-Synthesis Duality and WH Frames} 

WH frames are complete or overcomplete (i.e., redundant) WH sets with a certain guaranteed numerical stability of reconstruction 
\cite{groebook,Fei_Str_97,christensen03} (see the box \textsc{Weyl-Heisenberg Frames}). When decomposing a signal $x(t)$ into a WH frame 
$\{g_{n,k}(t)\}$ with dual WH frame $\{\gamma_{n,k}(t)\}$, we would first compute the expansion coefficients $\langle x,\gamma_{n,k} \rangle$ 
(analysis stage) and then reconstruct $x(t)$ according to $x(t)=\sum_{n\in\Z}\sum_{k\in\Z} \ist \langle x,\gamma_{n,k} \rangle g_{n,k}(t)$ (synthesis stage).
In OFDM systems, the transmitter performs \emph{synthesis} of the transmit signal according to
\revv{\eqref{eq:ofdmmod}}
with the data symbols $c_{n,k}$ playing the role of the expansion coefficients, and the receiver performs \emph{analysis} according to 
\revv{\eqref{eq:ofdmdemod}}.
This apparent duality is closely related to the duality and biorthogonality theory for WH frames \cite{wex90,janjf95,daubjf95}.

\begin{floatbox}[t!]
\vspace*{1mm}
\begin{center}
\begin{minipage}[t]{0.98\columnwidth}

\begin{center}
{\sc Weyl-Heisenberg Frames}
\end{center}
\vspace*{-1mm}

For $g(t) \!\in\! L_{2}(\R)$ and $T,F\!>\!0$, a function set $(g,T,F) = \{ g(t-nT)e^{j2\pi kFt} \}_{n,k\in\Z}$ is called a WH frame or Gabor frame for $L_2(\R)$ 
if for all $x(t)\in L_2(\R)$
\[
A\|x\|^{2}\,\le\,\sum_{n\in\Z} \sum_{k\in\Z} |\langle x,g_{n,k} \rangle|^{2}\,\le\,B\|x\|^{2} 
\]
with $0 \!<\! A \!\le\! B \!<\!  \infty$ \cite{groebook,Fei_Str_97,christensen03}. In what follows, we use the tightest constants $A$ and $B$; these are called 
lower and upper frame bound, respectively. The frame operator $\S$ is defined as the positive definite linear operator that maps $L_{2}(\R)$ onto $L_{2}(\R)$
according to
\[
({\S}x)(t)= \sum_{n\in\Z} \sum_{k\in\Z} \langle x,g_{n,k} \rangle g_{n,k}(t).
\]
For a WH frame $(g,T,F)$, the (minimal) {dual WH frame} is given by the set $(\gamma,T,F)$, where $\gamma(t)=({\S}^{-1}g)(t)$. 
The lower and upper frame bounds of the dual frame are given by $1/B$ and $1/A$, respectively. Using dual WH frames $(g,T,F)$ and $(\gamma,T,F)$, 
every signal $x(t) \in L_{2}(\R)$ can be decomposed as 
\begin{equation}
x(t) = \sum_{n\in\Z} \sum_{k\in\Z} \langle x,\gamma_{n,k} \rangle g_{n,k}(t)
= \sum_{n\in\Z} \sum_{k\in\Z} \langle x,g_{n,k} \rangle \gamma_{n,k}(t) .
\label{eq:frame_decomp}
\end{equation}

A WH frame is called tight if $A \!=\! B$. For a tight WH frame, we have ${\S}=A{\mathds I}$, where ${\mathds I}$ is the identity operator,
and hence $\gamma(t)=\frac{1}{A}g(t)$. If $(g,T,F)$ is a WH frame, $({\S}^{-1/2}g,T,F)$ is a tight WH frame with $A \!=\! B \!=\! 1$. 
Here, ${\S}^{-1/2}$ is the inverse positive definite square root of ${\S}$. 

In general, it is difficult to determine whether a given WH set $(g,T,F)$ is a WH frame. Intuitively, choosing $T$ and $F$ too large leaves ``gaps'' in $L_{2}(\R)$. 
Indeed, it can be shown that for $g(t) \in L_{2}(\R)$ and $TF>1$, the WH set $(g,T,F)$ cannot be a frame for $L_2(\R)$. 
The elements $g_{n,k}(t)$ of a WH frame with $TF=1$ are necessarily linearly independent, whereas WH frames with $TF<1$ necessarily 
have linearly dependent elements $g_{n,k}(t)$. 
Therefore, $(g,T,F)$ can be a frame for $L_2(\R)$ only if $TF \le 1$, i.e., when the TF grid
$\big\{(nT,kF)\big\}_{n,k\in\Z}$ is sufficiently dense. We note that WH analysis and synthesis can be interpreted as
the analysis and synthesis stage, respectively, of a DFT filter bank \cite{vaid93}.

\end{minipage}
\end{center}
\vspace*{-1mm}
\end{floatbox}

Duality and biorthogonality theory states that the WH sets $(g,T,F) \!=\! \{g(t \!-\! nT)e^{j2\pi kFt}\}_{n,k\in\Z}$ and $(\gamma,T,F)$
are biorthogonal if and only if the associated WH sets $(g,1/F,1/T)$ and $(\gamma,1/F,1/T)$ are dual frames; furthermore,
the WH set $(g,T,F)$ is orthogonal if and only if the associated WH set $(g,1/F,1/T)$ is a tight frame
(cf.\ the box \textsc{Duality and Biorthogonality}). The design of biorthogonal and orthogonal OFDM systems is therefore reduced to the 
widely studied problem of designing, respectively, dual and tight WH frames \cite{daub92}.

\begin{floatbox}[t!]
\vspace*{1mm}
\begin{center}
\begin{minipage}[t]{0.98\columnwidth}

\begin{center}
{\sc Duality and Biorthogonality}
\end{center}
\vspace*{-1mm}
In the finite-dimensional (cyclic) case, the duality and biorthogonality relation for WH frames essentially follows from the Poisson summation formula \cite{wex90}. 
We take all signals to be discrete-time and $N$-periodic and consider the WH frames $\{g_{n,k}[\ii]=g[\ii-nL]e^{j 2\pi\frac{k\ii}{M}}\}_{n\in \{0,\ldots,N/L-1\}, \, k\in \{0,\ldots,M-1\}}$
and $\{\gamma_{n,k}[\ii]=\gamma[\ii-nL]e^{j 2\pi \frac{k\ii}{M}} \}_{n\in \{0,\ldots,N/L-1\}, \, k\in \{0,\ldots,M-1\}}$ 
with time-shift parameter $L$ and frequency-shift parameter $1/M$, where $L,M\in \mathbb{N}$ and $M \ge L$. We assume that $N$ is an integer multiple of both $L$ and $M$.

We want to show that the WH frames $\{g_{n,k}[\ii]\}$ and $\{\gamma_{n,k}[\ii]\}$ are dual if and only if the WH sets 
$\{\tilde{g}_{n,k}[\ii]=g[\ii-nM]e^{j 2\pi \frac{k\ii}{L}} \}_{n\in \{0,\ldots,N/M-1\}, \, k\in \{0,\ldots,L-1\}}$ and
$\{\tilde{\gamma}_{n,k}[\ii]=\gamma[\ii-nM]e^{j 2\pi \frac{k\ii}{L}} \}_{n\in \{0,\ldots,N/M-1\}, \, k\in \{0,\ldots,L-1\}}$
with time-shift parameter $M$ and frequency-shift parameter $1/L$ are biorthogonal, i.e., 
$$
\langle \tilde{g}_{n,k},\tilde{\gamma}_{n^\prime,k^\prime}\rangle=\frac{L}{M} \, \delta_{n-n^\prime} \ist \delta_{k-k^\prime} \,.
$$

We start by noting that duality of $\{g_{n,k}[\ii]\}$ and $\{\gamma_{n,k}[\ii]\}$ (cf.\ \eqref{eq:frame_decomp}) is equivalent to the completeness relation
\begin{equation}
\sum_{n=0}^{N/L-1}\sum_{k=0}^{M-1} g_{n,k}[\ii]  \gamma^{\ast}_{n,k}[\ii^\prime] = \delta_{\ii-\ii^{\prime}} \, .
\label{eq:complete}
\end{equation}
The left-hand side of this relation can be shown to equal
\begin{equation}
M \! \sum_{n=-\infty}^{\infty} \!\Bigg[ \delta_{\ii-\ii^{\prime}-nM} 
\! \sum_{n'=0}^{N/L-1} \! f_{n}[\ii-n'L] \Bigg] \ist ,
\label{eq:lhs}
\end{equation}
where $f_n[\ii] = g[\ii]\gamma^{\ast}[\ii-nM]$. Furthermore, the Poisson summation formula yields
\begin{equation}
\sum_{n'=0}^{N/L-1} \! f_n[\ii-n'L]  
= \frac{1}{L}\sum_{k=0}^{L-1} F_n\bigg[\frac{kN}{L}\bigg]\,e^{j2\pi\frac{\ii k}{L}} ,
\label{eq:poisson}
\end{equation}
with $F_{n}[k]\,= \sum_{\ii=0}^{N-1} f_n[\ii]e^{-j2\pi\frac{k\ii}{N}}$.
Realizing that $F_n[kN/L]$\linebreak 
$= \langle\tilde{g},\tilde{\gamma}_{n,k}\rangle$,
and inserting into \eqref{eq:poisson} and, in turn, \eqref{eq:lhs}, we see that the left-hand side of \eqref{eq:complete} equals
\[
\frac{M}{L} \! \sum_{n=-\infty}^{\infty} \!\Bigg[ \delta_{\ii-\ii^{\prime}-nM} \sum_{k=0}^{L-1} \langle\tilde{g},\tilde{\gamma}_{n,k}\rangle \,e^{j2\pi\frac{\ii k}{L}} \Bigg] \ist .
\]
Thus, we can conclude that $\{g_{n,k}[i]\}$ and $\{\gamma_{n,k}[i]\}$ are dual if and only if $\langle \tilde{g},\tilde{\gamma}_{n,k}\rangle=\frac{L}{M} \, \delta_{n} \delta_{k}$,
i.e., if and only if the WH sets $\{\tilde{g}_{n,k}[i]\}$ and $\{\tilde{\gamma}_{n,k}[i]\}$ are biorthogonal.
\end{minipage}
\end{center}
\vspace*{-1mm}
\end{floatbox}

\vspace*{-1mm}

\subsection{Effect of a Doubly Dispersive Channel}

Consider an OFDM system with transmit pulse $g(t)$ and receive pulse $\gamma(t)$.
The transmit signal $x(t)$ is distorted by a TF dispersive channel $\H$ and contaminated by additive noise $w(t)$, 
resulting in the receive signal $y(t)=({\H}x)(t)+w(t)$. The OFDM demodulator output $\hat{c}_{n,k} =\langle y,\gamma_{n,k}\rangle$ then equals
\begin{equation}\label{eq:isiici_0}
\hat{c}_{n,k} \ist\ist=\, {H}_{n,k}\,c_{n,k} + I_{n,k} + w_{n,k} \,,
\end{equation}
where $H_{n,k}=\langle {\H}g_{n,k}, \gamma_{n,k} \rangle$ is the complex gain factor affecting the desired symbol $c_{n,k}$,
$I_{n,k}$ summarizes the interference caused by all other symbols $c_{n^{\prime}\!,k^{\prime}}$, $(n^{\prime}\!,k^{\prime}) \neq (n,k)$, and 
$w_{n,k}=\langle w,\gamma_{n,k} \rangle$. The interference term $I_{n,k}$ is given by
\begin{equation}\label{eq:isiici}
I_{n,k} \ist=\! 
\sum_{(n'\!,k')\,\in\, \Z^2\backslash (n,k)} \!\!\langle {\H}g_{n'\!,k'}, \gamma_{n,k} \rangle \,c_{n^{\prime}\!,k^{\prime}} \,.
\end{equation}
This comprises interference from symbols at different times $n'\rmv\neq n$ (intersymbol interference, ISI) and at different frequencies $k'\rmv\neq k$ 
(intercarrier interference, ICI).

ISI and ICI are negligible if $\H$ is underspread and $g(t)$, $\gamma(t)$, $T$, and $F$ are chosen appropriately as discussed in the next subsection.
In that case, the input-output relation \eqref{eq:isiici_0} decouples into a set of non-interfering parallel scalar channels according to
\begin{equation} \label{eq:diag}
\hat{c}_{n,k} \approx H_{n,k} \,c_{n,k} + w_{n,k} \,.
\end{equation}
This {approximate diagonalization of an underspread channel $\H$ drastically simplifies receiver tasks like data detection and channel estimation.
Note that $g_{n,k}(t)$ and $\gamma_{n,k}(t)$ can be viewed as {approximate singular functions} of $\H$ and $H_{n,k}=\langle {\H}g_{n,k}, \gamma_{n,k} \rangle$ 
as the corresponding approximate singular values. For normalized pulses $g(t)$ and $\gamma(t)$, it can be shown that $H_{n,k} \approx L_\H(nT,kF)$.
In the case of a time-invariant channel, CP-OFDM turns linear convolution into cyclic convolution. The corresponding channel matrix is 
circulant and diagonalized by the FFT (on which CP-OFDM is based) \cite{Peled80}, so that \eqref{eq:diag} becomes exact.

\vspace*{-1mm}

\subsection{Pulse Design} 

Next, we consider the problem of designing the transmit pulse $g(t)$ and the receive pulse $\gamma(t)$ such that 
small ISI and ICI are obtained \cite{boel_spie99,Matz-Charlypaper07,schniter_tsp07}. We note that OFDM systems with sophisticated ISI/ICI-reducing pulse shapes 
are currently hardly used in practice. This can be attributed to the fact that ISI and ICI can alternatively be mitigated
using equalization \cite{schniter_tsp07}. 

For WSSUS channels, the mean power of the interference term $I_{n,k}$ in \eqref{eq:isiici} can be shown to equal \cite{Matz-Charlypaper07}
\[
P_I = \mathsf{E}\{ |I_{n,k}|^2 \} = \iint C_\H(\tau,\nu)\,\tilde{A}_{g,\gamma}(\tau,\nu)\,d\tau\,d\nu ,
\vspace*{-1mm}
\]
with
\[
\tilde{A}_{g,\gamma}(\tau,\nu) \ist=\!  \sum_{ (m,l)\,\in\, \Z^2\backslash(0,0)} \!\! |A_{g,\gamma}(\tau-mT,\nu-lF)|^2, 
\]
where $A_{g,\gamma}(\tau,\nu) =\rmv \int g(t)\,\gamma^*(t \!-\! \tau)\,e^{-j2\pi\nu t}\ist dt 
=\langle g,\M_\nu\D_\tau \gamma\rangle$ is the cross-ambiguity function of $g(t)$ and $\gamma(t)$ \cite{fla-book2,groebook}. 
To obtain small ISI/ICI power $P_I$, the translates $|A_{g,\gamma}(\tau-mT,\nu-lF)|^2$, $(m,l) \in\Z\backslash (0,0)$,
should have little overlap with the channel's scattering function $C_\H(\tau,\nu)$. 
Clearly, making $P_I$ small by suitably choosing $g,\gamma,T$, and $F$ is easier for underspread channels with $C_\H(\tau,\nu)$ 
better concentrated around $(0,0)$. We then have to design 
pulses $g(t)$ and $\gamma(t)$ for which
$A_{g,\gamma}(\tau,\nu)$ decays rapidly, which in turn requires that the pulses be well TF localized \cite{Matz-Charlypaper07}.
An example of well-localized biorthogonal pulses is shown in Fig.~\ref{fig:pulses_ambifunc}. 
Biorthogonality implies $A_{g,\gamma}(mT,lF) \!=\! \delta_m \delta_l$ and indeed
the zeros of $A_{g,\gamma}(\tau,\nu)$ for $(\tau,\nu)=(mT,lF)$, $(m,l) \in\Z\backslash (0,0)$,
are clearly visible in Fig.~\ref{fig:pulses_ambifunc}(c). 
Further numerical results for pulse designs and the associated ISI/ICI levels are provided in \cite{Matz-Charlypaper07}.

\begin{figure}
\renewcommand\baselinestretch{1.1}\small\normalsize
\begin{center}
\noindent\includegraphics[width=\columnwidth]{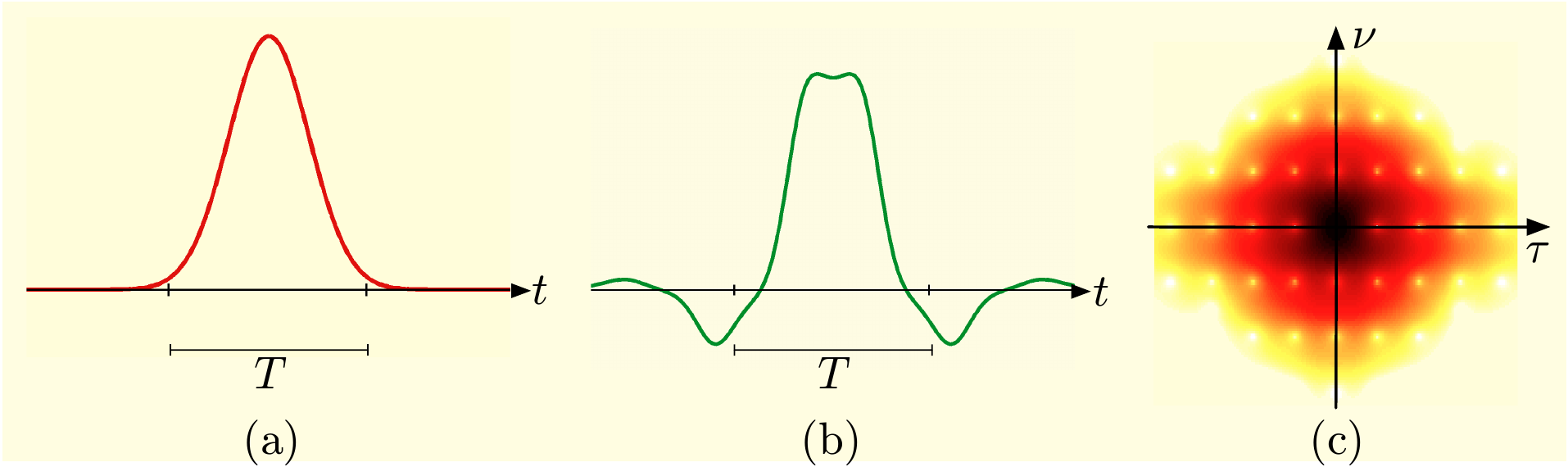}
\end{center}
\vspace*{-5mm}
\caption{\label{fig:pulses_ambifunc}
Example of a biorthogonal pulse-shaping OFDM system with $TF=1.33$, optimized for a TF dispersive channel with $\tau_{\max}=T/10$ and $\nu_{\max}=F/24$:
(a) transmit pulse $g(t)$, (b) receive pulse $\gamma(t)$, and (c) magnitude of the cross-ambiguity function $A_{g,\gamma}(\tau,\nu)$ (displayed on a log scale).}
\vspace*{-1mm}
\end{figure}

The analysis above shows that small ISI/ICI power $P_I$ requires an underspread channel $\H$ and pulses $g(t)$ and $\gamma(t)$ 
that are well localized both in time and in frequency. CP-OFDM employs a rectangular $g(t)$ (usually with a slight roll-off), whose
excellent time localization is well suited to purely time-dispersive channels; however, its poor frequency localization leads to ICI in frequency-dispersive channels.
Since well TF localized WH frames $(g,T,F)$ exist only for $TF \!<\! 1$ \cite{groebook}, it follows from duality and biorthogonality theory that well TF localized 
biorthogonal WH sets $(g,T,F)$ and $(\gamma,T,F)$ (which result in low ISI/ICI) exist only for $TF \!>\! 1$. Thus, there is a trade\-off between spectral efficiency and 
the amount of ISI/ICI incurred. Specifically, if $(g,T,F)$ with $TF \!=\! 1$ (maximal spectral efficiency) is an orthogonal basis for $L_{2}(\R)$, 
then $g(t)$ and its Fourier transform $G(f)$ cannot satisfy both $\int t^{2}|g(t)|^{2}\,dt \rmv<\rmv \infty$ and $\int f^{2}|G(f)|^{2}\,df \rmv<\rmv \infty$
\cite{groebook}.

\vspace{-1mm}

\section{Channel Capacity} 

We have seen that WH signal sets---corresponding to OFDM modulation---are well suited to communication over underspread TF dispersive channels since
they approximately diagonalize the channel. In addition, WH sets are also useful for characterizing the capacity limits of continuous-time WSSUS fading channels. 
We consider the noncoherent setting where neither transmitter nor receiver knows the channel realization but the transmitter knows the 
channel statistics (i.e., the scattering function $C_{\H}(\tau,\nu)$). The noncoherent capacity of fading channels \cite{gallager68,Ken69,bigshlom98} 
is the ultimate limit on the achievable rate since overhead transmissions like pilots and training sequences reduce spectral efficiency.

The standard approach to information-theoretic analyses of continuous-time channels is to discretize the input-output relation through a projection
onto the singular functions of the channel \cite{gallager68}. This yields a diagonalized discretized channel with noninteracting scalar input-output relations, 
similar to (\ref{eq:diag}). In the noncoherent case, this approach works only if all channel realizations have the same singular functions.
This is the case for time-invariant channels, where the eigenfunctions are complex sinusoids independently of the channel realization. 
However, for TF dispersive channels, the singular functions generally depend on the channel realization and do not have a specific structure.

Nevertheless, approximate capacity expressions can be obtained by using the channel discretization induced by OFDM 
\cite{durisiboel_book11}. We consider an underspread Gaussian WSSUS channel $\H$ with additive white Gaussian noise and 
OFDM modulation and demodulation using an orthonormal WH set $(g \!=\! \gamma,T,F)$, where $TF > 1$ and $g$ is well TF localized. 
An important advantage of using WH sets to discretize the channel (even though they do not diagonalize the channel exactly) is the fact that 
the channel coefficients $H_{n,k}$ in \eqref{eq:isiici_0} inherit the 2-D stationarity property of the continuous-time WSSUS channel.
In the low signal-to-noise ratio (SNR) regime, ignoring the ISI/ICI term $I_{n,k}$ (cf.\ \eqref{eq:isiici_0}) in the capacity computation leads to small 
approximation errors \cite{durisiboel_book11}. In the high-SNR regime, ISI/ICI cannot be neglected \cite{durisiboel_book11}. 

Using OFDM-based channel discretization, the capacity for low SNR satisfies \cite{durisiboel_book11}
$$
\mathsf{C}(\rho) \,\approx\, \mathsf{C}_{\mbox{\footnotesize AWGN}}(\rho) \ist-\rmv \iint \log\big(1+\rho\,C_{\H}(\tau,\nu)\big)\, d\tau\ist d\nu,
$$
where $\mathsf{C}_{\mbox{\footnotesize AWGN}}(\rho)$ is the capacity of a nondispersive additive white Gaussian noise channel
and the SNR $\rho$ is inversely proportional to the bandwidth. We see that $\mathsf{C}(\rho)$ is approximately equal to $\mathsf{C}_{\mbox{\footnotesize AWGN}}(\rho)$ 
minus a penalty term that is due to the unknown channel and increases with increasing channel spread (i.e., effective support of $C_{\H}(\tau,\nu)$). Furthermore,
$\mathsf{C}(\rho) \!\to\! 0$ as the bandwidth grows large. Intuitively, because of the uncorrelated scattering nature of the channel, the number of independent diversity
branches increases as the channel spread or the signal bandwidth increases and thus the receiver can no longer resolve the corresponding channel uncertainty. 
This also implies that $\mathsf{C}(\rho)$ has a maximum at a certain finite bandwidth. A detailed discussion of this phenomenon is provided in \cite{durisiboel_book11}.

For high SNR, $\mathsf{C}(\rho)$ is close to $\mathsf{C}_{\mbox{\footnotesize AWGN}}(\rho)$ for channel spreads occurring in wireless (radio) communications. 
Information-theoretic guidelines for the design of $(g,T,F)$ reveal that choosing $TF$ slightly larger than $1$ and using a root-raised-cosine pulse
for $g$ yields a lower bound on $\mathsf{C}(\rho)$ that is very close to the upper bound given by $\mathsf{C}_{\mbox{\footnotesize AWGN}}(\rho)$ \cite{durisiboel_book11}.

\section{System Identification}

The goal of channel/system identification \cite{Kai62} is to determine a channel/system $\H$ from the output signal $y(t)=(\H x)(t)$ 
given knowledge of the sounding (or probing) signal $x(t)$. This is relevant to dedicated channel sounding/measurement \cite{sounder_tc00}, 
channel estimation in the course of data transmission, and numerous other applications such as radar and sonar \cite{strohmer09}. 
Let us consider a TF dispersive channel $\H$ with spreading function $S_{\H}(\tau,\nu)$ supported in $[-\tau_{\max},\tau_{\max})\times [-\nu_{\max},\nu_{\max})$. 
In a practical scenario with finite input signal bandwidth $B$ and finite output signal observation time $D$, the input-output relation \eqref{eq:spreading_io} 
is discretized, resulting in an input-output relation of the form $\vect{y} \!=\! \A\s$ as explained in the box \textsc{Discretization}.

\begin{floatbox}[t!]
\vspace*{1mm}
\begin{center}
\begin{minipage}[t]{0.98\columnwidth}

\begin{center}
{\sc Discretization}  
\end{center}
\vspace*{-1mm}

We consider a transmit signal $x(t)$ that is band-limited to $[-B/2,B/2)$, and we observe the receive signal $y(t) = ({\H}x)(t)$ on $[-D/2,D/2)$.
Then, for $t\in[-D/2,D/2)$, the input-output relation \eqref{eq:spreading_io} becomes \cite{Bel63}
\begin{equation}
y(t) \approx \frac{1}{BD} \sum_{m\in \Z}\sum_{l \in \Z} S_\H\! \left(\frac{m}{B},\frac{l}{D} \right) \!\big(\M_{l/D}\D_{m/B} x)(t) .
\label{eq:sfunc_discrete}
\end{equation}
Thus, band-limiting the input and time-limiting the output leads to a discretization of \eqref{eq:spreading_io} with 
sample spacing $1/B$ and $1/D$ in delay and Doppler, respectively. 
For random (i.e., fading) channels, based on \eqref{eq:sfunc_discrete}, the concept of TF coherence regions and a TF rake receiver are developed in \cite{sayeed99}.

If the spreading function $S_{\H}(\tau,\nu)$ is supported in $[-\tau_{\max},\tau_{\max})\times [-\nu_{\max},\nu_{\max})$, 
only $|\mathcal{S}|=\lceil 4\tau_{\max}\nu_{\max}BD \rceil$ terms in \eqref{eq:sfunc_discrete} are non-zero. For $\nu_{\max} \rmv\rmv\ll\rmv\rmv B$,
the output signal $y(t)$ in \eqref{eq:sfunc_discrete} is approximately band-limited to $[-B/2,B/2)$. According to \cite{slepian76}, $y(t)$ restricted to $[-D/2,D/2)$
then lives in a signal space of dimension $N \!=\! \lceil BD \rceil$ that is spanned by an orthonormal basis of prolate spheroidal wave functions \cite{slepian76}.
Arranging the basis expansion coefficients of $y(t)$ in a vector $\vect{y}\in\C^{N}$, the input-output relation \eqref{eq:sfunc_discrete} translates into
\begin{equation}
\vect{y} = \A \s.
\label{eq:syseqyAs}
\end{equation}
Here, $\s\in \C^{|\mathcal{S}|}$ contains the $|\mathcal{S}|$ samples $S_{\H}(m/B,l/D)$, 
$(m/B,l/D) \in [-\tau_{\max},\tau_{\max})\times [-\nu_{\max},\nu_{\max})$,
and each column of $\A\in \C^{N \times |\mathcal{S}|}$ contains the expansion coefficients of a TF shifted version $\big(\M_{l/D}\D_{m/B} x)(t)$ of the input signal.

\end{minipage}
\end{center}
\vspace*{-1mm}
\end{floatbox}

The system identification problem thus amounts to reconstructing $\s$ from $\vect{y} \!=\! \A\s$, i.e., solving a linear system of equations. 
Clearly, for the existence of a unique solution $\s$, it is necessary that the number $|\mathcal{S}|$ of unknowns be smaller than or equal to the number $N$ 
of equations, which corresponds to the discrete underspread condition $|\mathcal{S}|\le N$. Due to $|\mathcal{S}| = \lceil 4\tau_{\max}\nu_{\max}BD \rceil$ 
and $N = \lceil BD \rceil$, this is equivalent to $\lceil 4\tau_{\max}\nu_{\max}BD \rceil \le \lceil BD \rceil$ and hence, effectively, to $d_\H=4\tau_{\max} \nu_{\max} \le 1$,
which implies that only underspread systems are identifiable. Sufficiency of the underspread condition $d_\H \le 1$ for identifiability is shown by 
explicitly constructing a sounding signal $x(t)$ such that ${\bf X}$ has full column rank. A viable choice for $x(t)$ is a (possibly weighted) Dirac train \cite{Kai62,sounder_tc00,kozek_identification_2005}. We have thus recovered the classical result by Kailath \cite{Kai62}, which states that a TF dispersive
system is identifiable if and only if it is underspread. Intuitively, in the overspread case, the system varies too fast 
to be identifiable. A generalized
version of Kailath's result was proven in \cite{kozek_identification_2005}. 
The results described above are non-parametric in that they do not impose structural assumptions on the system. Developing parametric equivalents using, e.g., 
the basis expansion model \cite{TsaGia96} is an interesting direction for further research.

The development above can be extended to systems whose spreading function support region is scattered across the delay-Doppler plane.
Such systems are identifiable if the overall support area of the spreading function is at most $1$ \cite{Bello69}. 
This result holds even if the spreading function support region is not known prior to identification \cite{heckelboel12}.}

It is commonly accepted that ``good'' sounding signals $x(t)$ have a rapidly decaying temporal autocorrelation function
(see, e.g., the references in \cite{sounder_tc00}). This statement specifically applies to time-invariant systems, which induce time shifts only. 
For TF dispersive systems, which cause both time and frequency shifts, our formulation of the identification problem shows that, for ${\bf X}$ in 
\eqref{eq:syseqyAs} to be well-conditioned, the TF translates of the sounding signal $x(t)$ should be as orthogonal to each other as possible.
This means that the auto-ambiguity function $A_{x,x}(\tau,\nu)$ should be small for $(\tau,\nu)=(m/B,l/D)$ with $(m,l) \in \Z^2\backslash (0,0)$.

\section{Conclusion} 

TF dispersive channels and WH function sets are central concepts in communications. Both are fundamentally based on the notion of TF shifts. 
Our aim in this paper was to demonstrate that the corresponding TF framework is not only conceptually interesting, but also provides powerful 
tools for solving problems such as pulse design in OFDM systems, characterization of the noncoherent capacity of continuous-time TF dispersive channels, 
and system identification and channel estimation. Furthermore, this TF framework applies in an almost one-to-one manner 
to other fields like radar and sonar (doubly spread targets \cite{vantrees3}) and quantum physics (quantization and coherent states \cite{folland89}). 
We thus hope that this paper will inspire innovative research and foster cross-fertilization between the signal processing, communications, information theory, 
physics, and mathematics communities.

\section*{Acknowledgments}
The authors are grateful to R.~Heckel for helpful comments and
to the authors of \cite{cfm_proc11} for permission to use the channel measurement
data in Fig.~\ref{fig:intro}. 
The work of F.~Hlawatsch and G.~Matz was supported by FWF Grants S10603 and S10606, respectively.


\section*{biographies}

\footnotesize

\noindent
\textit{\textbf{Gerald Matz}} (gmatz@nt.tuwien.ac.at) received the Dr.~techn.\ degree in electrical engineering 
from Vienna University of Technology, Austria, in 2000. 
He is Associate Professor with the Institute of Telecommunications, Vienna University of Technology. 
He coedited the book {\em Wireless Communications over Rapidly Time-Varying Channels} (New York: Academic, 2011). 
His research interests include signal processing, wireless communications, and information theory.
He serves on the IEEE SPS Technical Committees on Signal Processing for Communications and Networking
and on Signal Processing Theory and Methods. He currently is Associate Editor of the
{\sc IEEE Transactions on Information Theory} and was on the Editorial Board of
several signal processing journals.
He is a Senior Member of the IEEE. 
\vspace*{2mm}

\noindent
\textit{\textbf{Helmut B\"olcskei}} (boelcskei@nari.ee.ethz.ch) received the Ph.D.\ degree in electrical engineering from Vienna University of Technology, Vienna, Austria, in 1997. He has been with ETH Zurich since 2002, where he is Professor of Electrical Engineering.
He was in the founding teams of Iospan Wireless Inc.\ and Celestrius AG. His research interests are in information theory, mathematical signal processing, 
harmonic analysis, and statistics. He received the 2001 IEEE Signal Processing Society Young Author Best Paper Award, 
the 2006 IEEE Communications Society Leonard G. Abraham Best Paper Award, the 2010 Vodafone Innovations Award, 
a 2005 ETH ``Golden Owl'' Teaching Award, is a Fellow of the IEEE, and a 2011 EURASIP Fellow.
He was editor-in-chief of the {\sc IEEE Transactions on Information Theory}.

\vspace*{2mm}

\noindent
\textit{\textbf{Franz Hlawatsch}} (fhlawats@nt.tuwien.ac.at) received the Dr.\ techn.\ degree in electrical engineering from Vienna University of Technology, Austria, in 1988. Since 1983, he has been with the Institute of Telecommunications, Vienna University of Technology. 
His research interests include statistical signal processing, wireless communications, and sensor networks. 
He coedited the books {\em Time-Frequency Analysis: Concepts and Methods} (London: ISTE/Wiley, 2008) and
{\em Wireless Communications over Rapidly Time-Varying Channels} (New York: Academic, 2011). 
He was an associate editor of the {\sc IEEE Transactions on Signal Processing} and the {\sc IEEE Transactions on Information Theory}
and served on the IEEE SPS Technical Committee on Signal Processing for Communications and Networking.
He is a Fellow of the IEEE.

\end{document}